\documentclass[twocolumn,pre,showpacs]{revtex4}

\usepackage[latin1]{inputenc}
\usepackage{graphicx}

\newcommand{\Tr}{{\rm Tr\ }}
\newcommand{\etal}{{\it et al\ }}
\begin{document}

\title[Generalised interacting self-avoiding trails]{Generalised interacting self-avoiding trails on the square lattice: phase diagram and critical behaviour}

\author{D P  Foster}

\affiliation{Laboratoire de Physique Th\'eorique et Mod\'elisation
(CNRS UMR 8089), Universit\'e de Cergy-Pontoise, 2 ave A. Chauvin
95302 Cergy-Pontoise cedex, France}


\begin{abstract}
A generalised model for interacting self-avoiding trails on the square lattice is presented and studied using numerical transfer matrix methods. The model differentiates between on-site double visits corresponding to collisions, and crossings. Rigidity is also included in the model. The model includes the Nienhuis O(n=0) model and the interacting self-avoiding trail model as special cases. It is shown that the generic type of collapse found is the same as the pure interacting self-avoiding trail model. \end{abstract}


\maketitle


Lattice walk models have been used for many decades as coarse-grained models of polymers in solution\cite{vanderbook}. The quality of the solvent is modelled by introducing short-ranged interactions\cite{flory,degennes75}.
Whilst the interacting self-avoiding walk, with an attractive interaction associated with nearest-neighbour non-consecutively visited sites, is the most studied lattice walk model, models in which an attractive interaction is associated with multiply visited sites have also attracted attention. These models fall into two types: the Vertex-Interacting Self-Avoiding Walk (VISAW) model\cite{blotenienuis} and the Interacting Self-Avoiding Trails model\cite{massih75}. The VISAW is a bond-avoiding walk which is allowed to revisit sites, but is not allowed to intersect itself, whereas the ISAT is allowed to intersect at a lattice site. 

The VISAW is rather well understood, partly because it corresponds to the n=0 limit of the $O(n)$ model introduced by Blöte and Nienhuis \cite{blotenienuis}. The ISAT model, on the other hand, has given rise to many apparently contradictory results\cite{lyklema}. In recent work we presented evidence that the collapse transition in the ISAT trail has the same correlation length exponent, $\nu$, as the VISAW but a different entropic exponent $\gamma$\cite{F09}. The apparent contradiction with the results of Owczarek and Prellberg\cite{OP95} is resolved by the realisation that the relation $\nu=1/d_f$ between the $\nu$ exponent and the fractal dimension does not always hold\cite{FP03}.

An interesting question arises as to the generic nature of the ISAT point. If the parameter space is enlarged to enable one to extend the collapse point to a line, is the ISAT the generic type of collapse, or does one recover a more conventional $\theta$-type collapse? This question was recently examined on the triangular lattice\cite{DOP10}, where the construction of the lattice leads naturally to two different attractive interactions; on the triangular lattice one may revisit a site once or twice without visiting the same bond twice. In this study Doukas \etal\cite{DOP10} found that the ISAT collapse corresponded to a special point separating an isotropic collapse phase, a crystalline phase and the swollen phase. In this phase space the ISAT collapse extended into a line of $\theta$-points.

In this brief report we study a generalised interacting self-avoiding trail model on the square lattice to examine how the collapse varies as the model is moved from the standard interacting trail model. In our model this may be achieved either by differentiating between crossings and collisions, or by introducing a rigidity (weighting straight sections with respect to corners).


The model studied here is defined as follows: consider all random walks on the square lattice which do not visit any lattice bond more than once. Doubly visited sites may correspond to either crossings or ``collisions", crossings are assigned an attractive energy $-\varepsilon_x$, whilst collisions are assigned an attractive energy $-\varepsilon_c$. We also introduce a penalty $\varepsilon_s$ for straight sections. The partition function for the model is
\begin{equation}
{\cal Z}=\sum_{\rm walks} K^N\tau_x^{N_x}\tau_c^{N_c} p^{N_s},
\end{equation} 
where $K$ is the step fugacity, $\tau_{x/c}=\exp(\beta\varepsilon_{x/c})$, $p=\exp(-\beta\varepsilon_s)$, $N$ is the length of the walk, and $N_x$ is the number of crossings and $N_c$ is the number of collisions.

\begin{figure}
\begin{center}
\includegraphics[width=8cm]{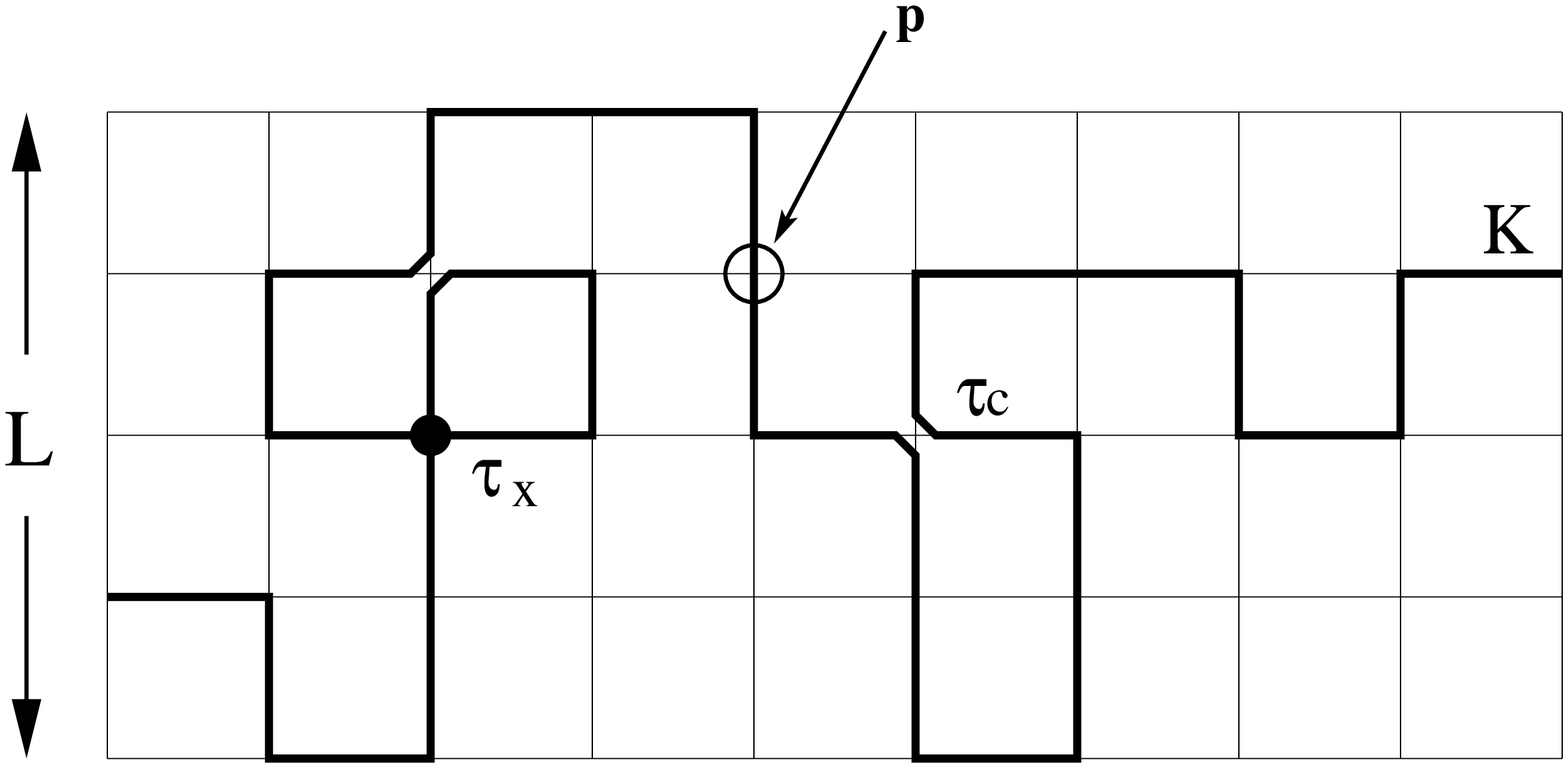}
\end{center}
\caption{The generalised interacting self-avoiding trail model showing the vertex crossings, weighted with a Boltzmann factor $\tau_x$, and the vertex collisions, weighted with a Boltzmann factor $\tau_c$. A weight $p$ is associated with straight segments and a fugacity $K$ controls the length of the walk. The walk is shown on a strip of width L = 5 and length $L_x=10$.}
\end{figure}

This partition function may be calculated exactly on a strip of length $L_x\to\infty$ and of finite width $L$ by defining a transfer matrix ${\cal T}$. If periodic boundary conditions are assumed in both directions, the partition function for the strip is given by:
\begin{equation}
{\cal Z}_L=\lim_{L_x\to\infty}\Tr\left({\cal T}^{L_x}\right).
\end{equation}
The free energy per lattice site, the density, and correlation length for the infinite strip may be calculated from the eigenvalues of the transfer matrix:
\begin{eqnarray}
f&=&\frac{1}{L}\ln\left(\lambda_0\right),\\
\rho(K,\tau)&=& \frac{K}{L\lambda_0}\frac{\partial \lambda_0}{\partial K},\\
\xi(K,\tau)&=&\left(\ln\left|\frac{\lambda_0}{\lambda_1}\right|\right)^{-1},
\end{eqnarray}
where $\lambda_0$ and $\lambda_1$ are the largest and second largest (in modulus) eigenvalues.

It is expected that ${\cal Z}$, $\rho$ and $\xi$ should have the following scaling forms close to the critical fugacity (for fixed $\tau$):
\begin{eqnarray}
{\cal Z}&\sim&|K-K_c|^{-\gamma},\\
\xi&\sim& |K-K_c|^{-\nu},\\\label{rs}
\rho_L(K)&=&\rho_\infty(K)+L^{1/\nu-2}\tilde{\rho}(|K-K_c|L^{1/\nu}).
\end{eqnarray}
${\cal Z}$ corresponds to the high temperature expansion of the susceptibility of an equivalent magnetic model, hence the use of the exponent $\gamma$. 

These scaling properties enable estimates of the critical lines to be calculated using a phenomenological renormalisation group method. For example a critical point estimate for a pair of lattice widths $L$ and $L^\prime$ is given by the solution of the equation\cite{mpn76}:
\begin{equation}\label{nrg}
\frac{\xi_L}{L}=\frac{\xi_{L^\prime}}{L^\prime}
\end{equation}
with estimates of the critical exponent $\nu$ given by:
 \begin{equation}\label{nuestim}
 \frac{1}{\nu_{L,L^\prime}}=\frac{\log\left(\frac{{\rm d}\xi_L}{{d}K}/\frac{{\rm d}\xi_{L^\prime}}{{d}K} \right)}{\log\left(L/L^\prime\right)}-1.
\end{equation}

For a more detailed discussion of the transfer matrix method, the reader is referred to the article of Blöte and Nienhuis~\cite{blotenienuis}.

\begin{figure}
\begin{center}
\includegraphics[width=8cm,clip]{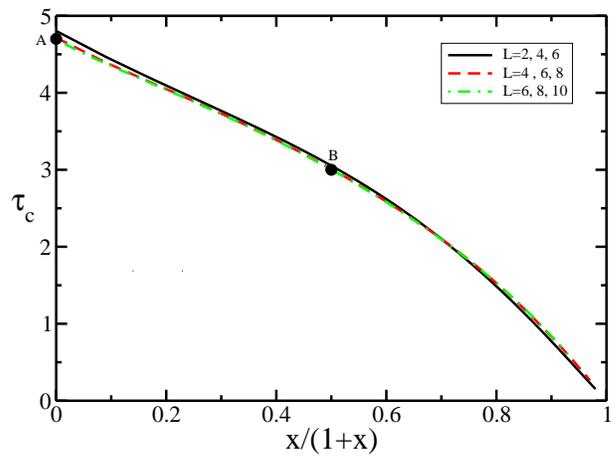}
\end{center}
\caption{(Color
 online) Phase diagram in the $x, \tau_c$ plane for the fully flexible generalised ISAT model ($p=1$). The point marked {\bf A} corresponds to the pure VISAW model and the point marked {\bf B} corresponds to the pure ISAT model.}\label{pd1}
\end{figure}

\begin{figure}
\begin{center}
\includegraphics[width=8cm,clip]{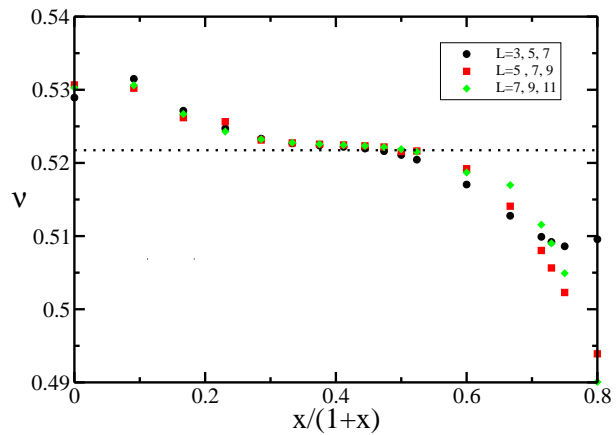}
\end{center}
\caption{(Color
 online) The $\nu$ exponent as a function of $\tau_c$ calculated for the flexible generalised ISAT model using the scaling of $\rho$ setting  $\lambda_1=\lambda_0=1$. The dotted line corresponds to $\nu=12/23$.}\label{nugenl1}
\end{figure}

\begin{figure}
\begin{center}

\includegraphics[width=8cm,clip]{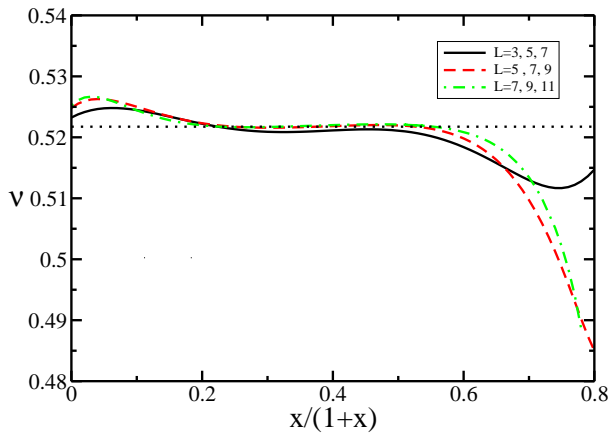}

\end{center}
\caption{(Color
 online) The $\nu$ exponent as a function of $\tau_c$ calculated for the flexible generalised ISAT model  using phenomenological RG (equations (\ref{nrg}) and (\ref{nuestim})). The dotted line corresponds to $\nu=12/23$.}\label{nugenrg}
\end{figure}


We first study the flexible generalised ISAT model. By flexible we mean that $p=1$ and the straight sections and corners are equally weighted. The phase diagram was calculated in the plane where the average length of the walk diverges, which occurs when $\lambda_1\to \lambda_0$, or equivalently when $\xi\to\infty$. This (critical) surface may then be estimated in two different ways - either by setting $\lambda_1=1$ (since $\lambda_0=1$), or by using the Phenomenological RG equation~(\ref{nrg}). In both cases the collapse transition may be found in a variety of ways, but commonly one looks at the variation of $\nu$ as $\tau_c$ (or $\tau_x$) is varied for fixed $x=\tau_x/\tau_c$. 
The crossings of $\nu$ give an 
estimate of the collapse transition. One may calculate $\nu$ either from the scaling behaviour of $\rho$ using (\ref{rs}), useful in the case we estimated the critical surface by setting $\lambda_0=\lambda_1=1$, or by the Phenomenological RG equation~(\ref{nuestim}). The two methods give superposable results for the phase diagram plotted in the critical surface, shown in Figure~\ref{pd1}. The estimates of $\nu$ calculated using the two methods are shown in Figures~\ref{nugenl1} and~\ref{nugenrg}.

Remembering that the VISAW model is expected to have $\nu=12/23$\cite{wbn}, and it is conjectured that the ISAT model also has $\nu=12/23$\cite{F09}, we may interpret the $\nu$ results as indicating that the entire collapse line to the left of the ISAT collapse point (B) is in the ISAT universality class, whilst at some stage the collapse transition becomes first order. It seems that the ISAT behaviour extends beyond $x=1$ ($x/(1+x)=0.5$). Both the VISAW model and the ISAT model are believed to have the same value of $\nu$, whilst the value of $\gamma$ was found to be different ($\gamma=53/46$ in the first case\cite{wbn} and conjectured to be $\gamma=22/23$ in the second\cite{F09}). This difference is similar to that seen between the $\theta$ point model on the regular lattice where $\gamma=8/7$\cite{ds} and on the Manhattan lattice $\gamma=6/7$\cite{bradley1990}. Both the Manhattan lattice $\theta$ point model and the ISAT model have the particularity that if the walks are grown dynamically, the construction of the walk can only fail if the walk meets its starting point. This feature is common to the whole collapse line except for the VISAW collapse point (A) ($x=0$); since crossings are forbidden at this point, when walks are grown they may find themselves trapped. This  suggests that the generic behaviour along the collapse line might be that of the ISAT, though this is not totally clear from the transfer matrix data available.

\begin{figure}
\begin{center}
\includegraphics[width=8cm,clip]{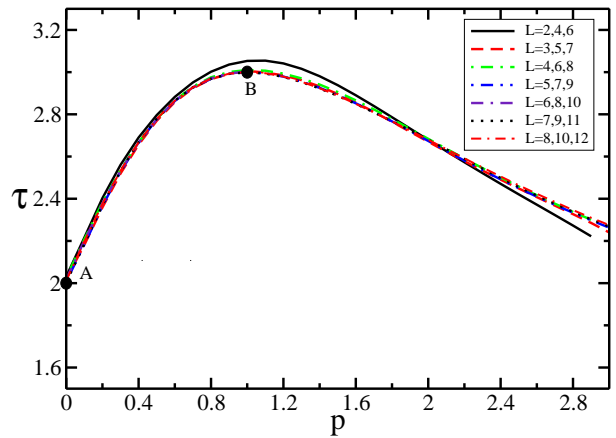}
\end{center}
\caption{(Color
 online) Phase diagram in the $p, \tau_c$ plain for the semi-flexible ISAT model ($x=1$). The point marked {\bf A} corresponds to the $O(n=0)$ Nienhuis-Blöte model with only corners. This point then maps onto the $\theta$ point on the Manhattan lattice or on the $L$ lattice (see text). The point marked {\bf B} corresponds to the pure ISAT model.}\label{pdtau_p}
\end{figure}

\begin{figure}
\begin{center}
\includegraphics[width=8cm,clip]{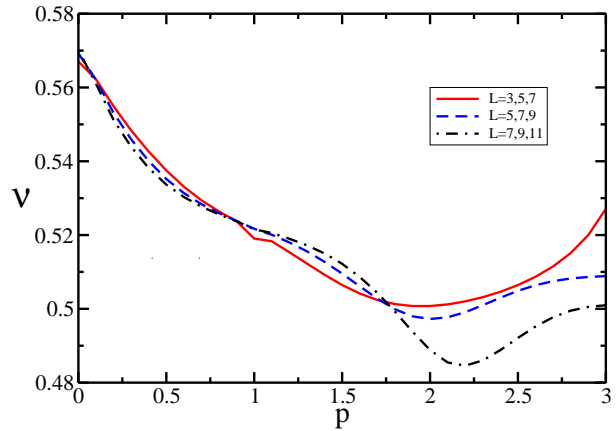}

\end{center}
\caption{(Color
 online) The $\nu$ exponent as a function of $\tau_c$ calculated 
using phenomenological RG (equations (\ref{nrg}) and (\ref{nuestim}))}\label{nu_p}
\end{figure}

On the square lattice, the other obvious means of moving away from the ISAT collapse is to include a rigidity to the walk. Here we chose to weight the straight sections relative to the corners with a weight $p$. This choice was made to coincide with the previous work of Blöte and Nienhuis\cite{blotenienuis}.

The phase diagram in the critical surface is calculated as before, and shown in Figure~\ref{pdtau_p} in the $\tau_c, p$ plane for $x=1$. 
The limit $p=0$ excludes the possibility of crossings, since a crossing necessarily requires two straight portions. At this point the model may be mapped into the interacting self-avoiding walk on the Manhattan lattice, which, as discussed above, is known to be $\theta$-like ($\nu=4/7$). Following from the results for the ISAT on the triangular lattice, where it was found that the $\theta$-type behaviour was more generic, we may expect that the collapse transition between (A) and the ISAT collapse point (B) might fall into a $\theta$-like universality class ($\theta$-like since the $\gamma$ exponent may differ from the $\theta$ point value $8/7$). The results, whilst not conclusive, do seem to indicate that the $\nu$ value calculated for the ISAT model extends into a plateau as the lattice widths increase (see Figure~\ref{nu_p}), leading to the conclusion that the ISAT behaviour is again in this case more generic. Again the results indicate that the plateau extends beyond the ISAT collapse point, before the collapse transition becomes first order for some value of $p>1$.

In this brief report we presented transfer matrix results for a generalised model for interacting self-avoiding trails in order to investigate the stability of ISAT collapse to varying the model parameters. On the triangular lattice Doukas \etal\cite{DOP10} presented convincing Monte Carlo results which showed that a collapse transition similar to the ISAT collapse was a special point in the phase diagram, and that generically the collapse transition was either in the $\theta$ universality class or first-order. The first type of collapse separates  a swollen walk from a dense isotropic phase and the second separates the swollen walk from an anisotropic crystalline collapsed phase. On the square lattice the ISAT collapse transition is an extended line separating the swollen phase from the crystalline phase. The collapse transition becomes first order when crossings dominate over collisions, which happens either for large enough $x=\tau_x/\tau_c$ or large enough $p$.


\begin{thebibliography}{12}
\bibitem{vanderbook} C. Vanderzande, \emph{Lattice Models of Polymers}, Cambridge: CUP, 1998
\bibitem{flory} P. Flory \emph{Principles of Polymer Chemistry}, Ithaca: Cornell University Press, 1971
\bibitem{degennes75} P.~G. de Gennes, \emph{J. Phys. Lett} {\bf 36} L55 (1975)
\bibitem{blotenienuis} H.~W.~J. {Bl\"ote} and B.~Nienhuis, \emph{J. Phys. A}, {\bf 22}  1415, (1989)
\bibitem{massih75} A.~R. Massih and M. A. Moore, \emph{J. Phys. A}, {\bf 8} 237, (1975)
\bibitem{lyklema} J. Lyklema, \emph{J. Phys. A}, {\bf18} L617 (1985); A. Guha, H.~A. Lim and Y. Shapir, \emph{J. Phys.}, {\bf A21} 1043, (1988); H. Meirovitch and H.~A. Lim, \emph{Phys. Rev. A}, {\bf 38} R1670 (1988); I. Chang and H. Meirovitch, \emph{Phys. Rev. Lett.}, {\bf 69}, 2232 (1992)
\bibitem{F09} D.~P. Foster,  \emph{J. Phys. A}, {\bf 42} 372002 (2009)
\bibitem{OP95} A.~L. Owczarek and T. Prellberg, \emph{J. Stat. Phys.} {\bf 79} 951 (1995)
\bibitem{FP03} D.~P. Foster and C. Pinettes \emph{J Phys A} {\bf 36} 10279 (2003)
\bibitem{DOP10} J Doukas, A.~L. Owczarek and T. Prellberg \emph{Phys Rev E} {\bf 82} 031103 (2010)
\bibitem{wbn} S.~O. Warnaar, M.~T. Batchelor, and B.~Nienhuis, \emph{J. Phys. A}, {\bf 25} 3077, (1992)
\bibitem{ds} B. Duplantier and H. Saleur, \emph{Phys. Rev. Lett.} {\bf 59} 539, (1987)
\bibitem{mpn76} Nightingale M P  1976 {\it Physica} {\bf A83} 561
\bibitem{bradley1990} R.~M. Bradley \emph{Phys. Rev. A} {\bf 41} 914 (1990)
\end{thebibliography}
\end{document}